\documentclass[conference]{IEEEtran}
\IEEEoverridecommandlockouts

\usepackage[english]{babel}
\usepackage{amsmath}
\usepackage{amsfonts}
\usepackage{calc}
\usepackage{subfigure}
\usepackage{tikz}
\usepackage{graphicx}
\usepackage[all]{xy}
\usepackage{array}
\usepackage{listings}
\usepackage{color}
\usepackage{longtable}
\usepackage{multirow}
\usepackage{rotating}
\usepackage[bookmarks=false]{hyperref}
\usepackage{textcomp}
\definecolor{listinggray}{gray}{0.9}
\definecolor{lbcolor}{rgb}{0.9,0.9,0.9}
\newcommand{\secref}[1]{Section~\ref{#1}}
\newcommand{\figref}[1]{Figure~\ref{#1}}
\newcommand{\lstref}[1]{Listing~\ref{#1}}
\newcommand{\tabref}[1]{Table~\ref{#1}}

\newcommand{\lstinl}[1]{\lstinline[basicstyle=\normalsize]{#1}}

\newcommand{\codechapel}[2]{\lstset{caption={#1},label=#2,language=c,morekeywords={%
proc, forall, in, var, reduce, scan, use}}}
\newcommand{\codecilk}[2]{\lstset{caption={#1},label=#2,language=c,morekeywords={%
spawn, sync, cilk, cilk_for, cilk_sync, cilk_spawn}}}

\newcommand{\codego}[2]{\lstset{caption={#1},label=#2,language=c,morekeywords={%
func, go, chan, bool, var}}}

\newcommand{\codetbb}[2]{\lstset{caption={#1},label=#2,language=c,otherkeywords={%
}}}

\newcommand{\chapinline}[1]{\codechapel{}{}\lstinline[basicstyle=\normalsize]{#1}}
\newcommand{\cilkinline}[1]{\codecilk{}{}\lstinline[basicstyle=\normalsize]{#1}}
\newcommand{\goinline}[1]{\codego{}{}\lstinline[basicstyle=\normalsize]{#1}}
\newcommand{\tbbinline}[1]{\codetbb{}{}\lstinline[basicstyle=\normalsize]{#1}}

\newcommand{\stext}[1]{\mbox{\footnotesize #1}}
\newcommand{\chapel}{\mbox{\footnotesize Chapel}}
\newcommand{\cilk}{\mbox{\footnotesize Cilk}}

\newcommand{\go}{\mbox{\footnotesize Go}}

\newcommand{\tbb}{\mbox{\footnotesize TBB}}

\newcommand{\rank}{\mathit{rating}}
\newcommand{\speedup}{\mathit{speedup}}

\begin{document}

\title{Benchmarking Usability and Performance of Multicore Languages}

\author{
\IEEEauthorblockN{Sebastian Nanz$^1$, Scott West$^1$, Kaue Soares da Silveira$^{2,\star}$\thanks{$^\star$ The ideas and opinions presented here are not necessarily shared by my employer.}, and Bertrand Meyer$^1$}
  \vspace*{1ex}
\IEEEauthorblockA{
  \begin{tabular}{cc}
    $^1$ ETH Zurich, Switzerland             & $^2$ Google Inc., Zurich, Switzerland \\
    \texttt{firstname.lastname@inf.ethz.ch}  & \texttt{kaue@google.com} \\
  \end{tabular}
  \vspace{2ex}
}
}


\maketitle

\begin{abstract}
Developers face a wide choice of programming languages and libraries supporting multicore computing. Ever more diverse paradigms for expressing parallelism and synchronization become available while their influence on usability and performance remains largely unclear.
This paper describes an experiment comparing four markedly different approaches to parallel programming: Chapel, Cilk, Go, and Threading Building Blocks (TBB).
Each language is used to implement sequential and parallel versions of six benchmark programs. The implementations are then reviewed by notable experts in the language, thereby obtaining reference versions for each language and benchmark. The resulting pool of 96 implementations is used to compare the languages with respect to source code size, coding time, execution time, and speedup.
The experiment uncovers strengths and weaknesses in all approaches, facilitating an informed selection of a language under a particular set of requirements. The expert review step furthermore highlights the importance of expert knowledge when using modern parallel programming approaches.
\end{abstract}


\section{Introduction}
\label{sec:introduction}

The industry-wide shift to multicore processors is expected to be permanent because of the physical constraints preventing frequency scaling~\cite{Asanovic:2009:VPC:1562764.1562783}. Since coming hardware generations will not make programs much faster unless they harness the available multicore processing power, parallel programming is gaining much importance.
At the same time, parallel programs are notoriously difficult to develop. On the one hand, concurrency makes programs prone to errors such as atomicity violations, data races, and deadlocks, which are hard to detect because of their nondeterministic nature. On the other hand, performance is a significant challenge, as scheduling and communication overheads or lock contention may lead to adverse effects, such as parallel slow down.

In response to these challenges, a plethora of advanced programming languages and libraries have been designed,  promising an improved development experience over traditional multithreaded programming, without compromising performance. These approaches are based on widely different programming abstractions. A lack of results that convincingly characterize both usability and performance of these approaches makes it difficult for developers to confidently choose among them. Current evaluations are typically based on classroom studies, e.g.\ \cite{Hochstein:2005:PPP:1105760.1105800, rossbach-et-al:2010:transactional, Cantonnet04productivityanalysis, Ebcioglu06experiment}; however, the use of novice programmers imposes serious obstacles to transferring experimental results into practice.

This paper presents an experiment to compare multicore languages and libraries, applied to four approaches: Chapel~\cite{Chamberlain:2007:PPC:1286120.1286123}, Cilk~\cite{Blumofe95cilk:an}, Go~\cite{golang}, and Threading Building Blocks (TBB)~\cite{DBLP:books/daglib/0018624}. 
The approaches are selected to cover a range of programming abstractions for parallelism, are all under current active development and backed by large corporations.
The experiment uses a two-step process for obtaining the program artifacts, avoiding problems of other comparison methodologies. First, an experienced developer implements both a sequential and a parallel version of a suite of six benchmark programs. Second, experts in the respective language review the implementations, leading to a set of reference versions. All experts participating in our experiment are high-profile, namely either leaders or prominent members of the respective compiler development teams. 
This process leads to a solution pool of 96 programs, i.e.\ six problems in four languages, each in a sequential and a parallel version, and before and after expert review. This pool is subjected to program metrics that relate both to usability (source code size and coding time) and performance (execution time and speedup). 
%
The experiment then statistically relates, for each metric, the approaches to each other. The results can thus provide guidance for choosing a suitable language according to usability and performance criteria. 
%

The remainder of this paper is structured as follows. \secref{sec:experimental-design} describes the experimental design of the study. \secref{sec:programming-models} provides an overview of the approaches chosen for the experiment. \secref{sec:results} discusses the results of the experiment. \secref{sec:related-work} presents related work and \secref{sec:conclusion} concludes with an outlook on future work.


\section{Experimental design}
\label{sec:experimental-design}

This section presents the research questions addressed by the study and the design of the experiment to answer them.

\subsection{Research questions}
\label{sec:research-questions}

Approaches to multicore programming are very diverse. To begin with, they are rooted in one basic programming paradigm such as imperative, functional, and object-oriented programming or multi-paradigmatic combinations of these. A further distinction is given by the communication paradigm used, such as shared memory and message-passing or their hybrids. Lastly, they differ in the programming abstractions chosen to express parallelism and synchronization, such as Fork-Join, Algorithmic Skeletons~\cite{Cole:1991:ASS:128874}, Communicating Sequential Processes (CSP)~\cite{Hoare:1978:CSP:359576.359585}, or Partitioned Global Address Space (PGAS)~\cite{DBLP:reference/parallel/X11lz} mechanisms. 

In spite of this diversity, all multicore languages share two common goals: to provide improved language \emph{usability} by offering advanced programming abstractions and run-time mechanisms, while facilitating the development of programs with a high level of \emph{performance}. While usabibility and performance are natural goals of any programming language, both aspects are particularly relevant in the case of languages for parallelism. First, achieving only average performance is not an option: parallel languages are employed precisely because of performance reasons. Second, usability is crucial because of the claim to be able to replace traditional threading models, which are ill-reputed precisely because of their lack of usability: unrestricted nondeterminism and the use of locks are among the aspects branded as error-prone. 

A comparative study of parallel languages has to evaluate both usability and performance aspects.
Hence, the abstract research questions are: 
\begin{description}
\setlength{\itemsep}{0.5ex}
\item[\textit{Usability}] \hspace{1ex} How easy is it to write a parallel program in language $L$?
\item[\textit{Performance}] \hspace{4ex} How efficient is a parallel program written in language $L$?
\end{description}
While correctness of the program is without doubt the most important dimension, it must be treated as a prerequisite to both the usability and performance evaluation if they are to make sense. Therefore we measure the usability of obtaining a program that solves a problem $P$ \emph{correctly}, and the performance of that program.

The abstract research questions need to be translated into concrete ones that correspond to measurable data. In this experiment, the following concrete research questions are investigated:
\begin{description}
\setlength{\itemsep}{0.5ex}
\item[\textit{Source code size}] \hspace{8ex} What is the size of the source code, measured in lines of code (LoC), of a solution to problem $P$ in language $L$?
\item[\textit{Coding time}] \hspace{4ex} How much time does it take to code a solution to problem $P$ in language $L$?
\item[\textit{Execution time}] \hspace{6ex} How much time does it take to execute a solution to problem $P$ in language $L$ if $n$ processor cores are available?
\item[\textit{Speedup}] \hspace{1ex} What is the speedup of a parallel solution to problem $P$ in language $L$ over the fastest known sequential solution to problem $P$ in language $L$ if $n$ processor cores are available?
\end{description}
We argue that both source code size and coding time strongly relate to usability. Shorter coding time can be due to many factors: availability of powerful language constructs that simplify the implementation; better support of the developer's reasoning, resulting in fewer iterations to obtain a correct program; availability of better documentation, examples, or development tools, speeding up the programmer's dwell time with a language issue; and others. While the measurement of coding time abstracts from such specific reasons, it is clear that it is directly related to these usability issues. 
While usability might also be captured by qualitative methods, e.g.\ asking about the perceived benefit of using a particular approach, we opted in this experiment to deal with quantifiable data only. Qualitative methods could however help to investigate further aspects of language usability, e.g.\ maintainability, which are not covered by our choice of measures.

Execution time and speedup relate to performance. Measuring speedup in addition to execution time gives important benefits: being a relative measure, it factors out performance deficiencies also present in the sequential base language, drawing attention to the power of the parallel mechanisms offered by the language; it also reveals scalability problems.

From the concrete research questions, it is apparent that the experimental setup has to provide for both a set $\mathbb{L}$ of languages as well as a set $\mathbb{P}$ of parallel programming problems. Sections~\ref{sec:selection} and~\ref{sec:benchmark-problems} explain how these sets were chosen.

\begin{table*}[htb]
  \centering
\def\arraystretch{1.3}
{\small
  \begin{tabular}{m{0.8cm}|m{5cm}|m{4.5cm}|m{3.5cm}|m{0.6cm}|m{1.5cm}}
Name   & Programming abstraction                  & Communication paradigm          & Programming paradigm & Year & Corporation \\
\hline
Chapel & Partitioned Global Address Space  & message passing / shared memory & object-oriented & 2006 & Cray Inc. \\
Cilk   & Structured Fork-Join                     & shared memory                   & imperative / object-oriented     & 1994 & Intel Corp. \\
Go     & Communicating Sequential Processes & message passing / shared memory & imperative      & 2009 & Google Inc. \\
TBB    & Algorithmic Skeletons                    & shared memory                   & C++ library         & 2006 & Intel Corp. \\
  \end{tabular}
}
  \vspace{2ex}
  \caption{Main language characteristics}
  \label{tab:languages}
\end{table*}

\subsection{Selection of languages}
\label{sec:selection}

The set $\mathbb{L}$ of languages considered in the experiment, further discussed in \secref{sec:programming-models}, was selected in the following manner. We first collected parallel programming approaches using web search and surveys, e.g.\ \cite{Skillicorn98modelsand}, resulting in 120 languages and libraries. We then applied two requirements to narrow down the approaches: 
\begin{itemize}
\setlength{\itemsep}{0.5ex}
\item The approach is under active development. This criterion was critical for the study, as it ensures that notable experts in the language are available during the expert review phase.
\item The approach has gained some popularity. This criterion was used to ensure that the results obtained by the experiment stay relevant for a longer time. Backing by a large corporation and a substantial user base were taken as signs of popularity.
\end{itemize}
The first criterion was by far the most important one, eliminating 73\% of approaches, as many languages were academic or industrial experiments that appeared to have been discontinued. Among the remaining approaches, about half of them were considered popular approaches. In the final selection, we preferred those approaches which would add to the variety of programming paradigms, communication paradigms, and/or programming abstractions considered. Well established approaches such as OpenMP and MPI (industry de-facto standards for shared memory and message passing parallelism) were not considered, as we wanted to focus on cutting-edge approaches.

\subsection{Benchmark problems}
\label{sec:benchmark-problems}

The set $\mathbb{P}$ of parallel programming problems was chosen from already suggested problem sets in the literature, e.g.~\cite{Wilson93assessingthe,Feo:1992:CSP:531027,Wilson95assessingand}. Reusing a tried and tested set has the benefit that estimates for the implementation complexity exist and that problem selection bias can be avoided by the experimenter.

After consideration, we chose the second version of the so-called Co\-wichan problems~\cite{Wilson95assessingand} (first version in \cite{Wilson93assessingthe}) as benchmarks, for the following reasons. First, the problems comprehend a wide range of parallel programming patterns, which is crucial to a comparative study. Second, given that we study four approaches in the experiment, it was important to keep the amount of time spent with every single problem reasonably small. The chosen problem set has been designed for this purpose~\cite{Wilson95assessingand}; in order to be more representative of large applications, the problems can however also be chained together.

In~\cite{Wilson95assessingand}, 13 such problems are presented. Again to keep the number of implementations manageable within the experiment, we selected the following six out of these: 
\begin{itemize}
\setlength{\itemsep}{0.5ex}
\item Random number generation (randmat)
\item Histogram thresholding (thresh)
\item Weighted point selection (winnow)
\item Outer product (outer)
\item Matrix-vector product (product)
\item Chaining of problems (chain)
\end{itemize}
Note that the last problem, chain, corresponds to a chaining together of the inputs and outputs of the other five. 


\subsection{Implementation}
\label{sec:implementation}

The benchmark problems were implemented in the chosen approaches observing the following practical considerations:
\begin{itemize}
\setlength{\itemsep}{0.5ex}
\item Use an experienced developer, but with no previous exposition to any of the chosen approaches, to implement all problems. 
\item Confirm that all solutions are correct by regression testing with predefined inputs and oracles.
\item Confirm that all parallel versions show some speedup over the respective sequential versions.
\item Use a version control system to measure the time to produce a solution 
through the commit times. Commit messages of the form 
\emph{``language-problem-variant keyword''} were used, 
where $\mbox{\emph{language}} \in \mathbb{L}$, 
$\mbox{\emph{problem}} \in \mathbb{P}$, 
$\mbox{\emph{variant}} \in \{\mbox{\emph{seq, par, expertseq,}}$ 
$\mbox{\emph{expertpar}}\}$, and 
$\mbox{\emph{keyword}} \in \{\mbox{\emph{start, pause, resume, done}}\}$.
\end{itemize}
We use an experienced developer (six years of experience; working at Google Inc.), rather than novice programmers, because of known issues with classroom approaches: parallel programs are hard to get right and to get to perform well, which makes the use of inexperienced programmers questionable. For example, Ebcioglu et al.~\cite{Ebcioglu06experiment} report that about one third of their students could not successfully complete a correct solution that achieved \emph{any} speedup, regardless of which of the three languages in their experiment was used; while this underlines the difficulty of parallel programming, it would be problematic to use such results to evaluate the languages.

Instead of a single experienced developer, one could also use a group of developers. The main problem with this approach is of a practical nature, in particular recruiting and budget. We decided instead to combine the use of a single developer with an expert review step (see \secref{sec:expert-feedback}), which has additional advantages.

\subsection{Expert review}
\label{sec:expert-feedback}

As a key step in the study, experts in the respective languages were asked to review the initial implementations. 
The expert review step has the advantage that it produces a set of reference versions of the benchmarks, creating a standard that holds across the different languages. 

The expert review had two main rounds. In the first round, the language designers or leaders of the respective development teams were contacted by email and asked for their help. Links to individual solutions in a browsable online repository were provided and brief instructions for the code review given, calling for any kind of feedback but especially on a) ways to make the implementations more concise or elegant and b) ways to improve their performance. In all cases, the initial contacts either provided comments themselves or recommended a member of their team, leading to the following list of experts:

\noindent
\begin{tabular}{@{}l@{~}p{7.75cm}}
\textit{Chapel} & Brad Chamberlain, Principal Engineer at Cray Inc. (technical lead on Chapel) \\[0.5ex]
\textit{Cilk} & Jim Sukha, Software Engineer at Intel Corp. (in the Cilk Plus development team, recommended by Charles E.\ Leiserson, one of the original Cilk designers) \\[0.5ex]
\textit{Go} & Luuk van Dijk, Software Engineer at Google Inc. (in the Go development team led by Andrew Gerrand, and recommended by him) \\[0.5ex]
\textit{TBB} & Arch D.\ Robison, Sr. Principal Engineer at Intel Corp. (chief architect of TBB) \\
\end{tabular}

After addressing comments from the first round, initial measurements were undertaken. The results were forwarded to the experts in a second round, together with links to improved implementations and requests for comments on the measurements. Comments from the second round were again incorporated.


\section{Languages}
\label{sec:programming-models}

This section provides the background on the approaches chosen for the
experiment: Chapel, Cilk, Go, and TBB. Table~\ref{tab:languages}
summarizes their characteristics, together with year of appearance,
and the corporation currently supporting further development.

\subsection{Chapel}
\label{sec:chapel}

Chapel~\cite{Chamberlain:2007:PPC:1286120.1286123} describes
parallelism in terms of independent computation implemented using
threads, but specified through higher-level abstractions. It won the
HPC Most Elegant Language award in 2011~\cite{hpcchallenge}. Chapel's
development effort to date has focused on correctness rather than
performance optimizations;\footnote{Personal communication with Brad
  Chamberlain.} expert comments for Chapel therefore reflect a tension
between writing for performance and writing for clarity.

We use parallel-for as a running example in
this section.
The concurrent variant of the \emph{for} statement,
\chapinline{forall}, provides parallel iteration over a range of
elements, as shown in \lstref{lst:chapel:parfor}; the loop will
execute, possibly in parallel, for all the elements between 1 and
\chapinline{n} (inclusive). Control continues with the statement
following the forall loop only after every iteration has been
evaluated.

\codechapel{Chapel: parallel-for}{lst:chapel:parfor}
\begin{lstlisting}
forall i in {1..n} {
  work (i);
}
\end{lstlisting}
\noindent The \chapinline{reduce} statement collapses a set of values down
to a summary value, e.g.\ computing the sum of the values in an
array. The \chapinline{scan} statement is similar to the reduction,
but stores the intermediate reductions in an array,
e.g.\ computing the prefix sum.

\subsection{Cilk}
\label{sec:cilk}

Cilk~\cite{Blumofe95cilk:an} exposes parallelism through high-level
primitives that are implemented by the runtime system, which takes
care of load balancing using dynamic scheduling through work
stealing. The language won HPC Best Overall Productivity award
2006~\cite{hpcchallenge}. Cilk's development started at MIT; since
2009 the technology has been further developed at Intel as Cilk Plus
(integration in commercial compiler, change of keywords, language
extensions).\footnote{Personal communication with Charles E.\
  Leiserson and Jim Sukha.}

The keyword \cilkinline{cilk_spawn} marks the concurrent variant of the
function call statement, which starts the (possibly) concurrent execution
of a function.  The synchronization statement \cilkinline{cilk_sync}
waits for the end of the execution of all the functions spawned in the
body of the current function; there is an implicit
\cilkinline{cilk_sync} statement at the end of all procedures.
Lastly, there is an additional \cilkinline{cilk_for} construct, see
\lstref{lst:cilk:parfor}.

\codecilk{Cilk: parallel-for}{lst:cilk:parfor}
\begin{lstlisting}
cilk_for (int i = 0; i < n; i++) {
  work (i);
}
\end{lstlisting}

\noindent This construct is a limited parallel variant of the normal
\cilkinline{for} statement, handling only simple loops.

\subsection{Go}
\label{sec:go}

Go~\cite{golang} is a general-purpose programming language targeted
towards systems programming. Parallelism is expressed using an
approach based on Communicating Sequential Processes
(CSP)~\cite{Hoare:1978:CSP:359576.359585}.

The statement \goinline{go} starts the execution of a function call as
an independent concurrent thread of control, or \emph{goroutine},
within the same address space. \emph{Channels} (indicated by the
\goinline{chan} type) provide a mechanism for two concurrently
executing functions to synchronize execution and communicate by
passing a value of a specified element type; channels can be
synchronous or asynchronous.

To construct a parallel-for loop, shown in \lstref{lst:go:parfor}, the
work gets dispatched to a channel (here \goinline{index}) from one go
routine, while \goinline{NP} goroutines fetch work from this channel
and process it.

\codego{Go: parallel-for}{lst:go:parfor}
\begin{lstlisting}
index := make(chan int)
done := make(chan bool)
NP := runtime.GOMAXPROCS(0)

go func() {
  for i := 0; i < n; i++ {
    index <- i
  }
  close(index)
}()

for i := 0; i < NP; i++ {
  go func() {
    for i := range index {
      work(i)
    }
    done <- true
  }()
}

for i := 0; i < NP; i++ {
  <-done
}
\end{lstlisting}

\noindent \goinline{NP} denotes the number of processors or threads
that are to be used.  To synchronize, each worker thread will send
\goinline{true} through a \mbox{\goinline{done}} channel
(\goinline{done <- true}); the main thread waits on this channel
(\goinline{<-done}) for \goinline{NP} values to come across the
channel before proceeding, indicating all workers have completed.

\subsection{Threading Building Blocks (TBB)}
\label{sec:tbb}

Threading Building Blocks (TBB)~\cite{DBLP:books/daglib/0018624} is a
parallel programming template library for the C++
language. Parallelism is expressed using Algorithmic
Skeletons~\cite{Cole:1991:ASS:128874}, and the runtime system takes
care of scheduling and load balancing using work stealing.

The function \tbbinline{parallel\_for} performs (possibly) parallel
iteration over a range of values, as shown in \lstref{lst:tbb:parfor};
the iteration is executed in non-deterministic order.

\codetbb{TBB: parallel-for}{lst:tbb:parfor}
\begin{lstlisting}
parallel_for(
  range(0, n),
  [=](range r) {
    for (size_t i = r.begin(); i != r.end(); i++) {
      work(i);
    }
  });
}
\end{lstlisting}

\noindent The \tbbinline{parallel_for} function takes as arguments the
range over which to iterate, and a lambda expression that itself will
be given a subrange that it may iterate across performing the work.

The \tbbinline{parallel_reduce} and \tbbinline{parallel_scan}
functions perform the same parallel operations as Chapel's
\chapinline{reduce} and \chapinline{scan}.


\newcommand{\spc}{\hspace{1ex}}
\newcommand{\ctr}[1]{#1~~}
\begin{table*}[htb]
  \centering
{\footnotesize\noindent
  \begin{tabular}[htb]{@{}l@{\spc}l@{\spc}||@{\spc}r@{\spc}r@{\spc}r@{\spc}r@{\spc}|@{\spc}r@{\spc}r@{\spc}r@{\spc}r@{\spc}|@{\spc}r@{\spc}r@{\spc}r@{\spc}r@{\spc}|@{\spc}r@{\spc}r@{\spc}r@{\spc}r@{\spc}|@{\spc}r@{\spc}r@{\spc}r@{\spc}r@{\spc}|@{\spc}r@{\spc}r@{\spc}r@{\spc}r@{\spc}}
  &Problem& \multicolumn{4}{c@{\spc}|@{\spc}}{randmat} & \multicolumn{4}{c@{\spc}|@{\spc}}{thresh} & \multicolumn{4}{c@{\spc}|@{\spc}}{winnow} & \multicolumn{4}{c@{\spc}|@{\spc}}{outer} & \multicolumn{4}{c@{\spc}|@{\spc}}{product} & \multicolumn{4}{c@{\spc}}{chain} \\
  &Version$^1$& \ctr{s} & ex-s & \ctr{p} & ex-p & \ctr{s} & ex-s & \ctr{p} & ex-p & \ctr{s} & ex-s & \ctr{p} & ex-p & \ctr{s} & ex-s & \ctr{p} & ex-p & \ctr{s} & ex-s & \ctr{p} & ex-p & \ctr{s} & ex-s & \ctr{p} & ex-p \\
  \hline\hline

\multirow{4}{*}{
  \begin{sideways}
    \begin{minipage}{1.3cm}
      \centering
      Source code size
    \end{minipage}
  \end{sideways}
} 
& Chapel & 33 & 32 & 33 & 32 & 58 & 51 & 58 & 61 & 71 & 61 & 72 & 74 & 55 & 58 & 55 & 58 & 34 & 31 & 34 & 36 & 145 & 130 & 145 & 159\\
& Cilk & 39 & 39 & 48 & 40 & 69 & 72 & 119 & 95 & 88 & 93 & 146 & 139 & 76 & 79 & 83 & 72 & 58 & 61 & 65 & 58 & 187 & 190 & 320 & 251\\
& Go & 37 & 54 & 52 & 71 & 73 & 77 & 141 & 118 & 116 & 112 & 144 & 191 & 88 & 78 & 103 & 98 & 74 & 68 & 89 & 86 & 204 & 160 & 345 & 330\\
& TBB & 38 & 38 & 52 & 53 & 69 & 69 & 110 & 98 & 78 & 78 & 142 & 137 & 72 & 69 & 83 & 81 & 49 & 50 & 63 & 62 & 172 & 171 & 302 & 302\\
\hline

\multirow{4}{*}{
  \begin{sideways}
  \begin{minipage}{1.3cm}
    \centering
    Coding time (min)
  \end{minipage}
  \end{sideways}
} 
& Chapel & 58 & 81 & 76 & 100 & 25 & 25 & 121 & 156 & 21 & 21 & 134 & 155 & 19 & 21 & 55 & 64 & 8 & 8 & 43 & 45 & 6 & 12 & 76 & 137 \\
& Cilk & 18 & 18 & 101 & 154 & 24 & 24 & 251 & 294 & 19 & 19 & 112 & 121 & 17 & 17 & 26 & 39 & 6 & 6 & 12 & 15 & 21 & 21 & 77 & 118 \\
& Go & 25 & 42 & 45 & 76 & 20 & 63 & 132 & 163 & 48 & 129 & 92 & 163 & 19 & 67 & 24 & 31 & 6 & 12 & 18 & 21 & 44 & 103 & 56 & 91 \\
& TBB & 15 & 15 & 35 & 37 & 31 & 31 & 196 & 207 & 25 & 25 & 41 & 43 & 28 & 28 & 32 & 43 & 15 & 15 & 23 & 23 & 12 & 12 & 24 & 26 \\
\hline

\multirow{4}{*}{
  \begin{sideways}
  \begin{minipage}{1.28cm}
    \centering
    Execution time (sec)$^2$
  \end{minipage}
  \end{sideways}
} 
& Chapel & 23.3 & 11.4 & 18.7 & 3.1 & 22.2 & 36.7 & 7.8 & 13.1 & 49.4 & 45.7 & 21.4 & 21.3 & 5.8 & 5.6 & 1.6 & 1.6 & 2.5 & 2.5 & 1.4 & 1.4 & 71.8 & 97.3 & 36.0 & 36.0 \\
& Cilk & 6.7 & 6.7 & 0.5 & 0.4 & 11.9 & 11.9 & 0.9 & 0.8 & 16.0 & 16.0 & 0.8 & 0.7 & 2.7 & 2.7 & 0.3 & 0.2 & 1.3 & 1.3 & 0.3 & 0.2 & 41.2 & 36.1 & 2.4 & 1.7 \\
& Go & 11.8 & 10.5 & 2.9 & 0.5 & 18.9 & 16.7 & 2.1 & 1.6 & 18.2 & 15.5 & 2.0 & 1.3 & 15.3 & 11.3 & 1.5 & 2.4 & 2.2 & 2.2 & 1.1 & 0.3 & 107.2 & 75.2 & 177.7 & 38.4 \\
& TBB & 5.3 & 5.3 & 0.3 & 0.2 & 9.3 & 9.4 & 1.2 & 0.6 & 9.7 & 9.7 & 1.0 & 1.0 & 1.9 & 1.9 & 0.3 & 0.3 & 1.3 & 1.3 & 0.2 & 0.2 & 35.2 & 35.5 & 2.8 & 2.8 \\
\hline

\multirow{4}{*}{
  \begin{sideways}
  \begin{minipage}{1.3cm}
    \centering
    Speedup$^2$
  \end{minipage}
  \end{sideways}
} 
& Chapel & - & - & 1.2 & 2.8 & - & - & 2.8 & 2.8 & - & - & 2.3 & 2.1 & - & - & 3.4 & 3.5 & - & - & 1.7 & 1.7 & - & - & 2.0 & 2.1 \\
& Cilk & - & - & 13.6 & 16.8 & - & - & 13.4 & 14.9 & - & - & 19.1 & 20.2 & - & - & 8.1 & 8.1 & - & - & 4.2 & 5.8 & - & - & 17.3 & 20.2 \\
& Go & - & - & 4.1 & 21.2 & - & - & 8.9 & 8.1 & - & - & 8.0 & 11.5 & - & - & 10.4 & 4.7 & - & - & 1.9 & 7.5 & - & - & 0.6 & 1.9 \\
& TBB & - & - & 20.7 & 21.2 & - & - & 8.1 & 14.8 & - & - & 9.4 & 9.5 & - & - & 7.4 & 7.4 & - & - & 7.2 & 7.3 & - & - & 12.5 & 12.6 \\
  \end{tabular}
}

{\footnotesize $^1$s: sequential; ex-s: expert-sequential; p: parallel; ex-p: expert-parallel}
{\footnotesize $^2$ average times and speedups are given \hfill}
  \vspace{2ex}
  \caption{Measurements for all metrics, across all languages, problems, and versions}
  \label{tab:data}
\end{table*}

\section{Results}
\label{sec:results}

This section presents and discusses the data collected in the experiment as defined in \secref{sec:experimental-design}. To facilitate replication of the results, an online repository\footnote{\url{https://bitbucket.org/nanzs/multicore-languages}} provides all the code as well as the analysis scripts. 

\subsection{Preliminaries}
\label{sec:preliminaries}

\tabref{tab:data} provides absolute numbers for all versions of the code, before and after expert review. 
Unless stated otherwise, the discussion of the data in \secref{sec:results} refers to the expert-parallel versions, i.e.\ the parallel versions obtained after expert review.

To facilitate comparison, all figures display the data in value-normalized form, namely relative to the smallest/fastest/etc. measurement per problem (which itself gets the value 1.0).

\subsubsection{Statistical evaluation} The results are statistically evaluated using the Wilcoxon signed-rank test (two-sided variant), a non-parametric test for paired samples. Specifically, for all metrics, each language is compared with each other language across all problems. We will say that ``$A$ is significantly different from $B$'' regarding a specific metric if $p < 0.05$; we will say that ``$A$ \emph{tends to be} different from $B$'' if $0.05 \leq p < 0.1$. 

We will represent the language relationships using graphs, where a solid arrow is drawn from $B$ to $A$ if $A$ is significantly better than $B$ in a certain metric; a dotted arrow is drawn if $A$ tends to be better than $B$. The ordering relations are transitive, but this will not explicitly be shown in the figures for clarity.

\subsubsection{Rating function} The statistical evaluation states the difference of two languages in qualitative terms, but does not expose the magnitude of this difference. The magnitude is important, however, because although two languages are significantly different regarding a certain metric, the relative difference might be small enough to be negligible in certain use cases. To address this, we define the \emph{average relative rating} of each language amongst the other languages, for a specific metric $m$:

\vspace{-2ex}
\begin{displaymath}
  \begin{array}{ll}
  \multicolumn{2}{c}{\displaystyle
  \rank_m(L) = \frac{1}{|\mathbb{P}|} \cdot \displaystyle \sum_{P \in \mathbb{P}} \dfrac{\displaystyle m(L, P)}{\displaystyle \min_{L' \in \mathbb{L}} m(L', P)}, \quad L \in \mathbb{L}}\\ \\
  \text{where:} \\
  \quad \mathbb{P} & \text{set of problems} \\
  \quad \mathbb{L} & \text{set of languages} \\
  \quad m: \mathbb{L} \times \mathbb{P} \rightarrow (0, \infty) & \text{metric function} \\
  \quad \rank: \mathbb{L} \rightarrow [1, \infty) & \text{rating function} \\
  \end{array}
  \label{eq:agg}
\end{displaymath}

For each language, the rating function calculates the average of the language's relative performance in each problem compared to the best performance of any language in the same problem. Thus, if the language was the best in all problems in a given metric, the result will be 1.0; a value of 2.0 for a given language and metric means that, on average, the language was 2 times ``worse'' (slower or larger, etc., depending on the metric) than the best language for that metric in each problem; and so on.

\subsection{Source code size}
\label{sec:source-code-size}

\begin{figure}[htbp]
  \centering
  \includegraphics[width=\columnwidth+0.3cm]{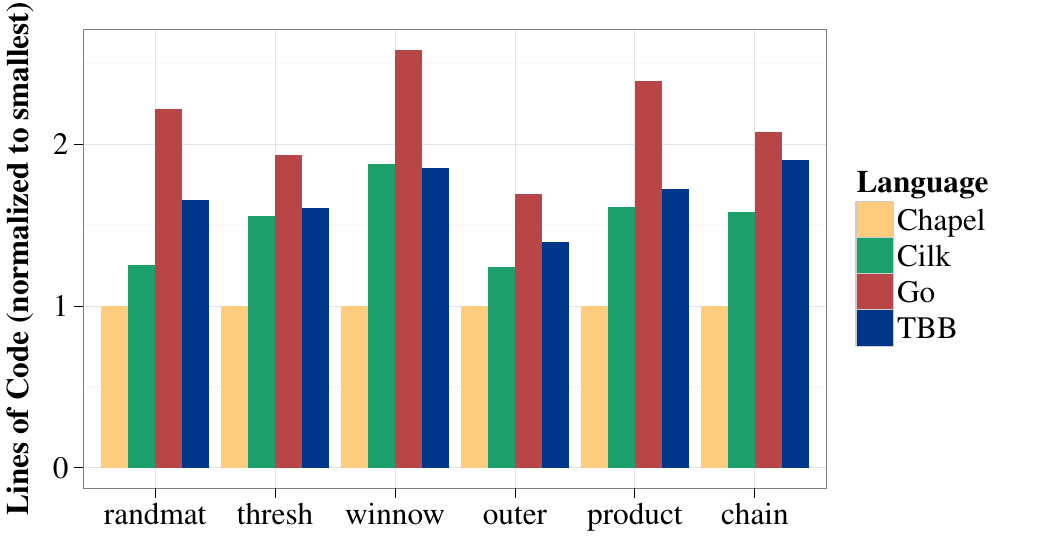}
  \caption{Source code size (LoC)}
  \label{fig:loc}
\end{figure}

The graph in \figref{fig:loc} shows the relative number of lines of source code (LoC) across all languages and problems, normalized to the smallest size in each problem.

Chapel shows the smallest code size in all of the problems, which can be explained by the conciseness of its language-integrated parallel directives. All the other languages are typically around 1.5-2.0 times larger, relative to Chapel's code size. Go's code size is the largest in all of the problems; reasons for this are the space taken for setting up the goroutines whenever a parallel operation is needed, and for synchronization with channels. Cilk and TBB hold a middle ground and are often comparable in code size.

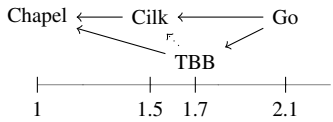
\begin{figure}[htbp]
  \centering
  \begin{tikzpicture}[scale=3]
    \node (chapel) at (1,0) {\chapel};
    \node (cilk)   at (1.5,0) {\cilk};
    \node (go)     at (2.1,0) {\go};
    \node (tbb)    at (1.7,-0.2) {\tbb};
    \path[->] (go) edge (cilk);
    \path[->] (go) edge (tbb);
    \path[->] (tbb) edge[dotted] (cilk);
    \path[->] (tbb) edge (chapel);
    \path[->] (cilk) edge (chapel);

    \draw[very thin,color=gray,step=0.3] (1,-0.3) grid (2.3,-0.3);
    \foreach \pos in {1, 1.5, 1.7, 2.1}
      \draw[shift={(\pos,-0.3)}] (0pt,1pt) -- (0pt,-1pt) node[below] {\stext{\pos}};
  \end{tikzpicture}
  \vspace{-2ex}
  \caption{Source code size: statistical ordering and rating}
  \label{fig:ord:size}
\end{figure}

Results of the Wilcoxon test and of the rating function are combined in \figref{fig:ord:size} (\secref{sec:preliminaries} explains how to interpret the graph). The placement of a language along the x-axis reflects its rating according to the rating function.

This confirms statistically the visual interpretation of \figref{fig:loc}. Chapel provides the most concise code overall, and Go the largest code size (on average about 2.1 times as large as Chapel's code). Cilk and TBB are in between, with Cilk tending to have smaller code sizes than TBB.

\subsection{Coding time}
\label{sec:coding-time}

\begin{figure}[htbp]
  \centering
  \includegraphics[width=\columnwidth+0.3cm]{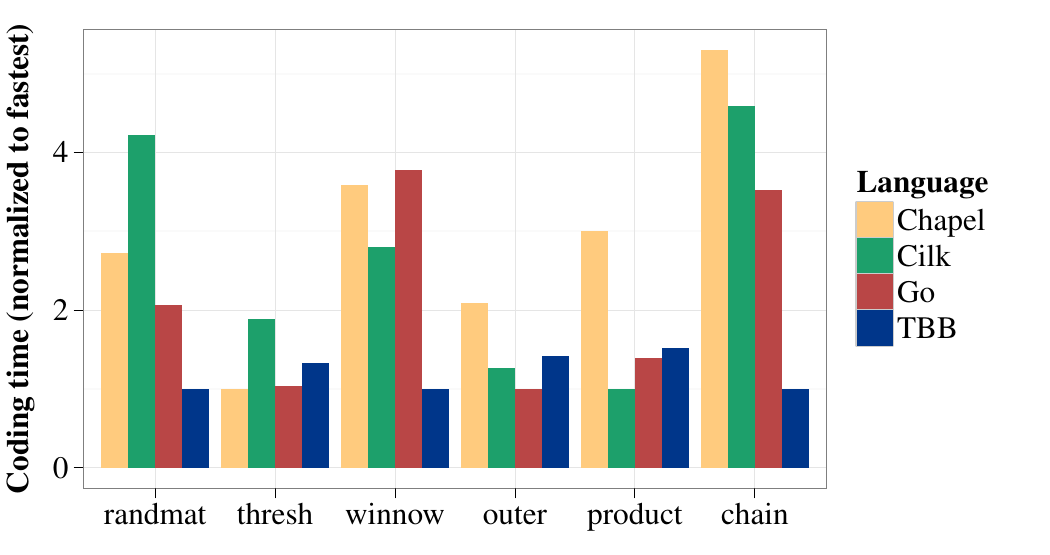}
  \caption{Coding time}
  \label{fig:time}
\end{figure}

\figref{fig:time} shows the relative time to code for each problem in each language. Note that times are cumulative in the following way: the coding time in the reference versions (expert-parallel time) is the sum of the time used to obtain the initial parallel version (parallel time) plus the time needed to refine it after the expert comments; as the parallel versions were based on the sequential ones, the parallel coding time itself includes the sequential coding time (sequential and expert-sequential time, respectively) in all cases. 

\begin{figure}[htbp]
  \centering
  \begin{tikzpicture}[scale=3]
    \node (chapel) at (3.0,0) {\chapel};
    \node (cilk)   at (2.6,0) {\cilk};
    \node (go)     at (2.1,0) {\go};
    \node (tbb)    at (1.2,0) {\tbb};

    \draw[very thin,color=gray,step=0.1] (1,-0.1) grid (3.2,-0.1);
    \foreach \pos in {1, 1.2, 2.1, 2.6, 3.0}
      \draw[shift={(\pos,-0.1)}] (0pt,1pt) -- (0pt,-1pt) node[below] {\stext{\pos}};
  \end{tikzpicture}
  \vspace{-2ex}
  \caption{Coding time: statistical ordering and rating}
  \label{fig:ord:codingtime}
\end{figure}
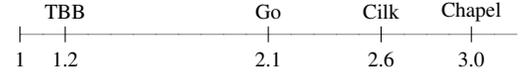

In contrast to the lines of code metric, the figure does not suggest any immediate conclusions. No clear ordering is visible, although TBB seems to have consistently low (but not always lowest) coding times. This is confirmed by the statistical evaluation, which yields no significant differences (displayed again as graph in \figref{fig:ord:codingtime}). The individual ratings show a clearer picture: coding in TBB takes on average only 1.2 times longer to code than the other three approaches, placing it at the top; Go, Cilk, and Chapel take on average at least 2.1 times longer, with Chapel taking 3.0 times longer. 

\subsection{Execution time}
\label{sec:execution-time}

\subsubsection{Measurement} The performance tests were run on a 4 $\times$ Intel Xeon Processor E7-4830 (2.13 GHz, 8 cores; total 32 physical cores) server with 256 GB of RAM, running Red Hat Enterprise Linux Server release 6.3. Language and compiler versions used were: chapel-1.6.0 with gcc-4.4.6, for Chapel; Intel C++ Compiler XE 13.0 for Linux, for both Cilk and TBB; go-1.0.3, for Go.

Each performance test was repeated 30 times, and the mean of the results was taken. All tests use the same inputs, the size-dominant of which is a $4 \cdot 10^4 \times 4 \cdot 10^4$ matrix (about 12 GB of RAM). This size, which is the largest input size all languages could handle, was chosen to test scalability. The language Go provided the tightest constraint, while the other languages would have been able to scale to even larger sizes. 

An important factor in the measurement is that for all problems the I/O time is significant, since they involve reading/writing matrices to/from the disk. In order for the measurements to not be dominated by I/O, a special flag \lstinl{is_bench} was added to every solution. This flag means that neither input nor output should occur and that the input matrices should be generated on-the-fly instead. All performance tests were run with the \lstinl{is_bench} flag set.

\subsubsection{Observations} 
\begin{figure}[ht]
  \centering
  \includegraphics[width=\columnwidth+0.3cm]{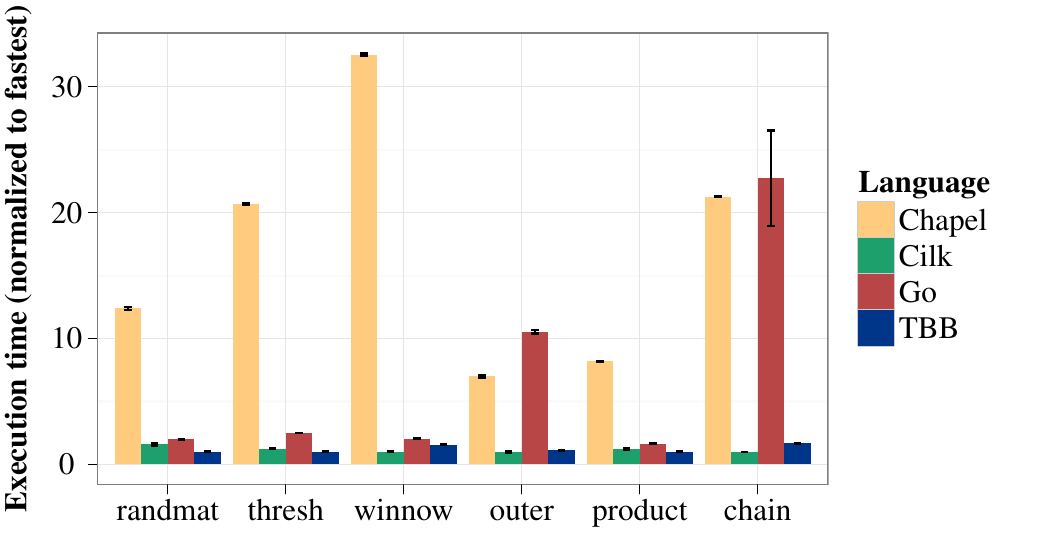}
  \caption{Execution time}
  \label{fig:exec:time}
\end{figure}

\figref{fig:exec:time} shows the relative execution time on 32 cores for each language and problem. The error bars show the 99.9\% confidence interval for the mean.

Chapel took the most time to execute in almost all problems. As mentioned in \secref{sec:chapel}, this reflects that the Chapel compiler is missing important optimizations, which were deferred to give first more attention to correctness. Also, all Chapel variables are default initialized; in particular the large matrix in the experiment will be zeroed, causing additional delay.  The distance to the other languages decreases significantly as the input size is decreased, hinting at the fact that the main problem is scalability (Chapel's speedup reaches a plateau early, as discussed in \secref{sec:speedup}).

Go shows uneven execution times across the problems, which might be explained by the language's lack of maturity (only 3 years old); the performance might show more stable results in the future. 
In particular, the execution time for the chain and outer problems are 
much higher than expected, 
they should be on the same order of magnitude as the other subproblems.
Chain additionally has a much higher variance than expected.

TBB and Cilk show consistently low execution times.

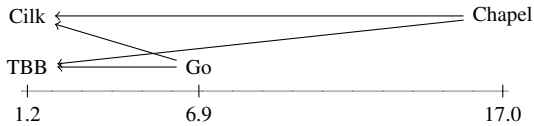
\begin{figure}[htbp]
  \centering
  \begin{tikzpicture}[scale=0.4]
    \node (chapel) at (17.0,1.5) {\chapel};
    \node (cilk)   at (1.2,1.5) {\cilk};
    \node (go)     at (6.9,-0.2) {\go};
    \node (tbb)    at (1.2,-0.2) {\tbb};

    \path[->] (chapel) edge (cilk);
    \path[->] (chapel) edge (tbb);
    \path[->] (go) edge (cilk);
    \path[->] (go) edge (tbb);

    \draw[very thin,color=gray,step=1] (1,-1) grid (17.2,-1);
    \foreach \pos in {1.2, 6.9, 17.0}
      \draw[shift={(\pos,-1)}] (0pt,6pt) -- (0pt,-6pt) node[below] {\stext{\pos}};
  \end{tikzpicture}
  \vspace{-2ex}
  \caption{Execution time: statistical ordering and rating}
  \label{fig:ord:exectime}
\end{figure}

This impression is confirmed statistically, as shown in \figref{fig:ord:exectime}. Both Chapel and Go exhibit a significantly slower execution time than Cilk and TBB. Considering the rating, TBB and Cilk are on par with a score of 1.2, followed by Go at 6.9 and Chapel at 17.0.

\subsection{Speedup}
\label{sec:speedup}

\begin{figure*}[htb]
  \centering
  \includegraphics[width=\textwidth-1.5cm]{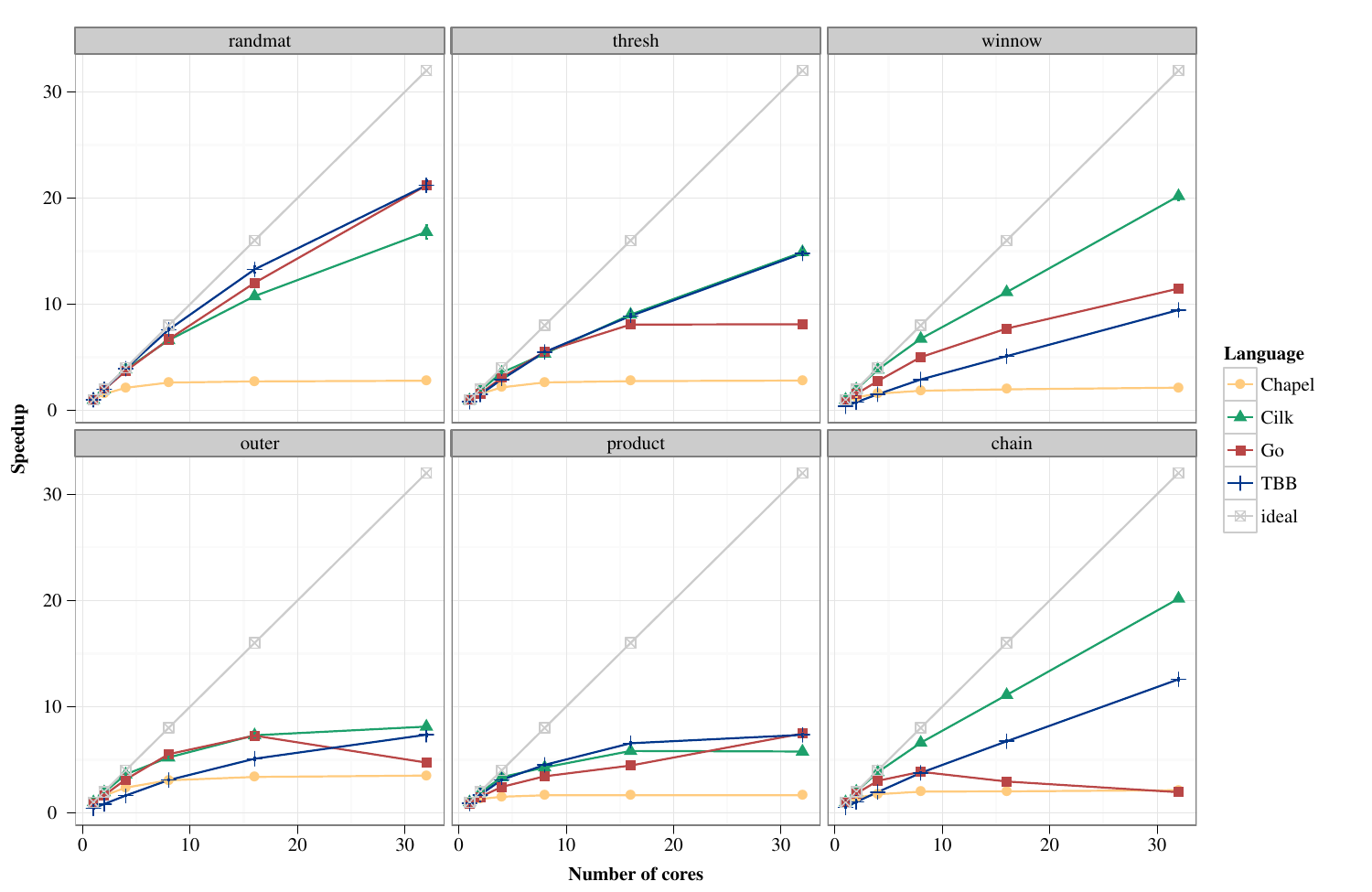}
  \caption{Speedup per problem}
  \label{fig:speedup:problem}
\end{figure*}

\subsubsection{Measurement} Speedup was measured across 1, 2, 4, 8, 16, and 32 cores, with respect to the \emph{fastest single thread implementation} in the respective language; this is the fastest implementation when executed on a single logical thread, i.e.\ either the sequential version itself, or the parallel version restricted to run on a single thread. 
\begin{displaymath}
  \begin{array}{lp{7cm}}
  \multicolumn{2}{c}{\displaystyle
  \speedup(n) = \frac{T_s}{T_p(n)}} \\ \\
  \text{where:} \\
  \quad T_s         & \text{fastest single thread implementation (sequen-} \text{\ \ \ \ tial or parallel)} \\
  \quad T_p(n)      & \text{execution time of the expert-parallel version} \text{\ \ \ \ with $n$ cores} \\
  \quad \speedup(n) & \text{speedup with $n$ cores} \\
  \end{array}
\end{displaymath}
%

\subsubsection{Observations} \figref{fig:speedup:problem} shows the speedup graphs per problem. The values are accurate within a 99.9\% confidence interval 
(error bars would not be visible on the plot).

For the problems product and outer, the speedup in all languages tends to plateau starting from about 16 cores. This can be partly attributed to the fact that the sequential versions already take very little time to execute on these problems; the input size would have to be further increased (but cannot without losing the ability to compare amongst all approaches, as discussed above). In all other problems, at least one language shows good scalability; as the number of cores increase, the speedup lines fan out, showing that there are significant differences.

Cilk and TBB show good scalability on these problems, with speedups of about 15--20 and 10--21 on 32 cores, respectively. 

Go's scalability is more uneven: in product and randmat it keeps up with the top performers; a plateau is visible in thresh at 16 cores; and speedup deterioration is detected in chain and outer. The deterioration might be caused by excessive creation of goroutines, generating scheduling and communication overheads. 

Chapel's speedup consistently plateaus early from around 4-8 cores and at a speedup of around 2-3 in all problems, but does not specifically underperform in any of them. This shows the need of an an overall improvement in the Chapel compiler's implementation (see discussion in \secref{sec:chapel}).

\begin{figure}[htbp]
  \centering
  \begin{tikzpicture}[scale=1.4]
    \node (chapel) at (6.4,0.3) {\chapel};
    \node (cilk)   at (1.0,0.3) {\cilk};
    \node (go)     at (2.9,-0.3) {\go};
    \node (tbb)    at (1.4,-0.3) {\tbb};

    \path[->] (chapel) edge (cilk);
    \path[->] (chapel) edge (tbb);
    \path[->] (chapel) edge[dotted] (go);
    \path[->] (go) edge[dotted] (tbb);

    \draw[very thin,color=gray,step=0.5] (1,-0.5) grid (6.6,-0.5);
    \foreach \pos in {2.9, 6.4}
      \draw[shift={(\pos,-0.5)}] (0pt,2pt) -- (0pt,-2pt) node[below] {\stext{\pos}};
    \foreach \pos in {1.0}
      \draw[shift={(\pos,-0.5)}] (0pt,2pt) -- (0pt,-2pt) node[below] {\stext{1.1}};    
    \foreach \pos in {1.4}
      \draw[shift={(\pos,-0.5)}] (0pt,2pt) -- (0pt,-2pt) node[below] {\stext{1.3}};    
  \end{tikzpicture}
  \vspace{-2ex}
  \caption{(Inverse of) speedup: statistical ordering and rating}
  \label{fig:ord:speedup}
\end{figure}
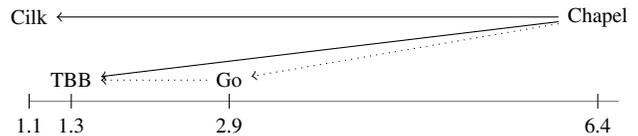

\figref{fig:ord:speedup} shows the results of the statistical tests and the application of the rating function, for the speedup at 32 cores. We opted for using the speedup at 32 cores, as it represents the best approximation available to the asymptotic speedup. Note that the rating function has to be modified slightly: since in the speedup measure larger is better, the inverse of the metric value is used.

Confirming the expectation from the speedup graphs, Chapel shows significantly worse speedup than Cilk and TBB, and tends to show worse speedup than Go. Cilk and TBB show no significant difference, while Go tends to show worse speedup than TBB.

\subsection{Lessons learned from the expert review}
\label{sec:influence-expert}

In the previous sections, the results were computed from the study's reference versions of the benchmark (i.e.\ after expert review). However, numerous lessons can also be drawn from the experts' suggestions on how to \emph{change} the original implementations. For space reasons, we present only two examples of such change requests (for Cilk and Go); a more complete analysis is available in a related publication~\cite{nanz-et-al:2013:examining}.



\codecilk{}{}
After expert review, all Cilk solutions decreased in source code size by up to about 20\%. This change can be traced back to one of the expert comments to replace \lstinline[basicstyle=\normalsize]{cilk_spawn}/\lstinline[basicstyle=\normalsize]{cilk_sync} style code (see \lstref{lst:cilk:spawnsync}), an idiom that Cilk has been known for originally.

\codecilk{Cilk: divide-and-conquer}{lst:cilk:spawnsync}
\begin{lstlisting}
void do_work(int begin, int end) {
  int middle = begin + (end - begin) / 2;
  if (begin + 1 == end) {
    work (begin);
  } else {
    cilk_spawn do_work (begin, middle);
    cilk_spawn do_work (middle, end);
  }
  cilk_sync;
}
. . .
cilk_spawn do_work (0, n);
\end{lstlisting}

The expert suggested to use \mbox{\lstinline[basicstyle=\normalsize]{cilk_for}} (\lstref{lst:cilk:parfor}) as it simplifies the code while doing the same recursive divide-and-conquer underneath, and should therefore be preferred.

Strong execution time improvements (up to about 80\% decrease) after expert review were visible for Go in many of the problems. This can be attributed to a single change in the way parallelism was achieved. In the non-expert versions, a divide-and-conquer pattern of the form displayed in \lstref{lst:go:divideandconquer} was used. Instead, the expert recommended the distribute-work-synchronize pattern of \lstref{lst:go:parfor}. While the divide-and-conquer approach creates one goroutine per row of the matrix, the distribute-work-synchronize approach creates one for each processor core; for large matrices, the overhead of the excessive creation of goroutines then causes a performance hit.


\codego{Go: divide-and-conquer}{lst:go:divideandconquer}
\begin{lstlisting}
func do_work(begin, end, done chan bool) {
  if (begin + 1 == end) {
    work (begin, done)
  } else {
    middle := begin + (end - begin) / 2
    go do_work (begin, middle, done)
    do_work (middle, end, done)
  }
}
. . .
done := make(chan bool)
go do_work(0, n, done)

for i := 0; i < nrows; i++ {
  <-done
}
\end{lstlisting}




\subsection{Threats to validity}
\label{sec:threats-to-validity}


As a threat to external validity, it is arguable whether the study results transfer to large applications, due to the size of the programs used. The modest problem size is intrinsic to the study: the use of top experts is crucial to reliably answer the research questions and, unfortunately, this also means that the program size has to remain reasonable to fit within the review time the experts were able to donate. However, a recent study~\cite{okur:2012:libraries} confirms that the amount of code dedicated to parallel constructs for 10K and 100K LOC programs is between 12 and 60 lines of code on average; this makes our study programs representative of the parallel portions of larger programs. 

Furthermore, only one developer is used to provide the base versions for all languages. While this would be a serious objection in other contexts, this a lesser threat to validity in our experiment. This is because the experiment builds on the concept of single \emph{reference} benchmark programs rather than groups of average-quality programs. This is indeed one of the innovations of our experimental setup, and avoids problems with using non-expert study participants, which are discussed in \secref{sec:related-work}. The base versions were only provided to enable the expert review step; while alternatively, we could have asked experts to implement the benchmark problems themselves from scratch, this would certainly have exceeded the time the experts were able to donate for the study. 

On the other hand, the choice of using expert-checked reference benchmark programs may threaten external validity in the sense that the results only hold for developers which are highly skilled in the approaches. Also, the influence of a learning effect when a single developer solves the same problem in different languages remains as a threat, which could be mitigated by using a group of developers.

Problem selection bias, a threat to internal validity, is avoided in part by using an existing problem set, instead of creating a new one. The threat that specific problems could be better suited to some languages than others remains, as it could already be present in the existing problem set. As a positive point, none of the experts criticized the choice of problems for evaluating their language.

Since the languages are based on very different fundamental designs, it is not immediately clear whether or not they are actually comparable. But as long as it is possible to solve a problem in all languages the metric comparison seems to be meaningful. Again, the experts were aware of the competing languages and did not challenge the choice of languages.
  


\subsection{Discussion}
\label{sec:discussion}

Considering all four metrics together (see the summary of the ratings in \tabref{tab:overallranks}), it becomes apparent that all four languages have individual strengths and weaknesses.

\begin{table}[htb]
  \centering
\def\arraystretch{1.1}
{\footnotesize
  \begin{tabular}{l|cccc}
         & Source       & Coding       & Execution    & (Inverse of) \\
         & code size    & time         & time         & speedup\\
\hline
Chapel   & \textbf{1.0} & 3.0          & 17.0         & 6.4 \\
Cilk     & 1.5          & 2.6          & \textbf{1.2} & \textbf{1.1} \\
Go       & 2.1          & 2.1          & 6.9          & 2.9 \\
TBB      & 1.7          & \textbf{1.2} & \textbf{1.2} & 1.3 \\
  \end{tabular}
}
  \vspace{2ex}
  \caption{Ratings (smaller is better; best in bold)}
  \label{tab:overallranks}
  \vspace{-2ex}
\end{table}

Chapel has incorporated parallel directives at the language level. This has an apparent advantage as the code size is consistently the smallest of all problems: there is a clear benefit to having language-level support for high-level operations.  However, the performance rates quite low, though this does not appear to be an inherent property of the language, but rather that the focus of the compiler implementation has been on other issues (see discussion in \secref{sec:chapel}).

Cilk's initial claim to fame was in lightweight tasks, which could quickly be balanced among many threads. Consequently, the language shows very strong performance results. Since then Cilk has gained a new keyword (\cilkinline{cilk_for}), giving it also some advantage on the code size metric.

Go has been designed for more general forms of concurrency; for example, using channels for communication but allowing shared memory where necessary is very flexible. Consequently, Go does not have extensive support for structured parallel computations, such as fork-join or parallel-for. This is evident in the source code size, which is often the largest. Go does an acceptable job on the performance measures, although some problems have been detected. Since the language is the youngest in the study (it appeared in 2009), the compiler is expected to mature in this respect.

TBB has no language-level support, it is strictly a library approach. However the library it provides is the most comprehensive of the four languages, containing algorithmic skeletons, task groups, synchronization and message passing facilities. The high level parallel algorithms were sufficient to implement every task in the benchmark set without dropping down to lower level primitives such as manual task creation and synchronization. TBB provides together with Cilk the best performance. Being a library for a well known language, it also has the fastest coding times.

Although none of the languages have any mechanisms to ensure freedom from concurrency issues such as data races or deadlocks, their common aim is to provide the ability to use built-in functionality to make the common cases easy and as safe as they can be.


\section{Related work}
\label{sec:related-work}

A number of related works present studies comparing approaches to parallel programming, albeit for languages different from the ones used in our experiment.

Szafron and Schaeffer~\cite{Szafron94experimentallyassessing} assess the usability of two parallel programming systems (a message passing library and a high-level parallel programming system) using a population of 15 students, and a single problem (transitive closure). Six metrics were evaluated: number of work hours, lines of code, number of sessions, number of compilations, number of runs, and execution time. They conclude that the high-level system is more usable overall, although the library is superior in some of the metrics; this highlights the difficulty in reconciling the results of different metrics. 
In contrast to this approach, we report no overall rank; instead we provide ranks within a metric, as the suitability of a language may depend on external factors that give different weight to each of the metrics.

Hochstein et al.~\cite{Hochstein:2005:PPP:1105760.1105800} provide a case study of the parallel programmer productivity of novice parallel programmers. The authors consider two problems (game of life and grid of resistors) and two programming models (MPI and OpenMP). They investigate speedup, code expansion factor, time to completion, and cost per line of code, concluding that MPI requires more effort than OpenMP overall in terms of time and lines of code.
%
%
Rossbach et al.~\cite{rossbach-et-al:2010:transactional} conducted a study with 237 undergraduate students implementing the same program with locks, monitors, and transactions. While the students felt on average that programming with locks was easier than programming with transactions, the transactional memory implementations had the fewest errors.
%
%
Ebcioglu et al.~\cite{Ebcioglu06experiment} measure the productivity of three parallel programming languages (MPI, UPC, and X10), using 27 students, and a single problem (Smith-Waterman local sequence matching). For each of the languages, about a third of the students could not achieve any speedup.
The methodology used in our experiment, namely using an experienced programmer and expert feedback, was able to avoid low-quality solutions. 

Nanz et al.~\cite{nanz-et-al:2011:design} present an empirical study with 67 students to compare the ease of use (program understanding, debugging, and writing) of two concurrency programming approaches (SCOOP and multi-threaded Java). They use self-study to avoid teaching bias and standard evaluation techniques to avoid subjectivity in the evaluation of the answers. They conclude that SCOOP is easier to use than multi-threaded Java regarding program understanding and debugging, and equivalent regarding program writing.
Pankratius et al.~\cite{Pankratius:2012:CFI:2337223.2337238} compare the languages Scala and Java using 13 students and one software engineer working on three different projects. The resulting programs are compared with respect to programmer effort, code compactness, language usage, program performance, and programmer satisfaction. They conclude that Scala's functional style does lead to more compact code and comparable performance.
Burkhart et al.~\cite{burkhart-et-al:2012:run} compare Chapel against non-PGAS models (Java Concurrency, OpenMP, MPI, CUDA, PATUS) in a classroom setting both in terms of productivity (working hours, parallel overhead, lines of code, learning curve) and performance. Results for Chapel were favorable on the productivity metrics, but lagged behind other languages on the performance side.

Cantonnet et al.~\cite{Cantonnet04productivityanalysis} analyze the productivity of two languages (UPC and MPI), using the metrics of lines of code and conceptual complexity (number of function calls, parameters, etc.), obtaining results in favor of UPC.
%
%
Bal~\cite{Bal91acomparative} is a practical study based on actual programming experience with five languages (SR, Emerald, Parlog, Linda and Orca) and two problems (traveling salesman, all pairs shortest paths). It reports the authors' experience while implementing the solutions.

It is worth noting that all the above studies either use novices as study participants (problem with ensuring a high quality of the code), or use implementations of the study authors (problem with experimenter bias, if the authors are also the designers of the approaches). In our experiment, we use reference implementations of benchmark programs, obtained by review with high-profile experts, avoiding these problems. 

\section{Conclusions}
\label{sec:conclusion}

We presented an experiment comparing four popular approaches to parallel programming, providing two main contributions. 
First, we defined a methodology for comparing multicore languages, involving reference implementations of benchmark programs, obtained through review by notable experts. We found that this methodology provides robust results as it ensures consistently high-quality program artifacts. 
Second, applying the experiment to Chapel, Cilk, Go, and TBB provided a detailed comparative study of usability and performance. The discussion of the differences of the languages was supported both by statistical tests and a rating function that quantifies these differences. This provided an unambiguous characterization of the approaches that can help developers choose among them.

Our methodology can serve as a template for further comparisons, and we plan to apply it to more languages in the future, e.g.\ to include widely used approaches such as OpenMP and MPI. The example of Go showed that there is also a need for benchmarks to evaluate languages with respect to more general concurrency patterns. Defining such a set of benchmarks is another interesting direction of future work. 

\section*{Acknowledgements}
We are grateful to Brad Chamberlain, Arch D.\ Robison, Jim Sukha, and Luuk van Dijk for devoting their time, effort, and expertise to this study.
This work was partially supported by the European Research Council under the European Union's Seventh Framework Programme (ERC Grant agreement no. 291389), FAPERGS/CNPq 008/2009 Research Project: ``GREEN-GRID: Sustainable High Performance Computing'', the Hasler Foundation, and ETH (ETHIIRA).


\bibliographystyle{IEEEtranS}
\bibliography{bibfile}

\end{document}